# Volatile evolution and atmospheres of Trans-Neptunian Objects

Leslie A. Young[a]

Felipe Braga-Ribas[b]

Robert E. Johnson[c]

[a] *Southwest Research Institute, Boulder CO, USA*
[b] *Federal University of Technology - Paraná/Brazil - UTFPR/Curitiba*
[c] *University of Virginia, Charlottesville, VA, USA*

**Abstract**

At 30-50 K, the temperatures typical for surfaces in the Kuiper Belt (e.g. Stern & Trafton 2008), only seven species have sublimation pressures higher than 1 nbar (Fray & Schmitt 2009): Ne, $N_2$, CO, Ar, $O_2$, $CH_4$, and Kr. Of these, $N_2$, CO, and $CH_4$ have been detected or inferred on the surfaces of Trans-Neptunian Objects (TNOs). The presence of tenuous atmospheres above these volatile ices depends on the sublimation pressures, which are very sensitive to the composition, temperatures, and mixing states of the volatile ices. Therefore, the retention of volatiles on a TNO is related to its formation environment and thermal history. The surface volatiles may be transported via seasonally varying atmospheres and their condensation might be responsible for the high surface albedos of some of these bodies. The most sensitive searches for tenuous atmospheres are made by the method of stellar occultation, which have been vital for the study of the atmospheres of Triton and Pluto, and has to-date placed upper limits on the atmospheres of 11 other bodies. The recent release of the Gaia astrometric catalog has led to a "golden age" in the ability to predict TNO occultations in order to increase the observational data base.

*Keywords:* Trans Neptunian objects, Kuiper Belt Objects, ices, atmospheres

## 1. Introduction

Several bodies in the Transneptunian region have volatiles on their surfaces that have significant vapor pressures at the temperatures of the outer solar system: $CH_4$, $N_2$, and CO (with CO only detected on Pluto and Triton). When present, volatiles may raise significant atmospheres around these bodies at some times during the orbit of these Trans-Neptunian Objects (TNOs). These would be vapor-pressure supported atmospheres, where the main atmospheric species exists also as a surface ice, and the surface pressure is a very sensitive function of that ice's temperature. Mars, Pluto, Triton, and, to some extent, Io are examples of vapor-pressure supported atmospheres. If, as expected, Eris, Makemake, and other TNOs are occasionally in this class at some time in their orbit, then vapor-pressure supported atmospheres would be more numerous than terrestrial atmospheres (Venus, Earth, Titan), or gas giants.

Rather than an oddity, vapor-pressure supported atmospheres may be the most common style of atmosphere in our solar system. These atmospheres are characterized by large seasonal pressure variations, and global transport of volatiles across the surface.

Vapor-pressure supported atmospheres may have been important in the evolution of the outer solar system. All the TNOs should have formed with some measure of these species, but not all TNOs have volatiles detected on their surface. The explanation is tied to the gaseous phase through atmospheric escape.

The atmospheres of Pluto and Triton have been extensively studied, by spacecraft, ground-based occultations and spectroscopy, and modeling. Great effort has been made to search for atmospheres around other TNOs. To date, only upper limits have been placed.

## 2. Spectral evidence of $N_2$, CO, and $CH_4$ on the surfaces of TNOs

Methane ice has been reported or inferred on the surfaces Triton, Pluto, Eris, Makemake, Quaoar, Varuna, Sedna, and 2007 $OR_{10}$. CO and $N_2$ ices have been directly detected on Pluto and Triton. $N_2$ ice has been inferred from its effect on the $CH_4$-dominated spectra of Eris and Makemake, and possibly Quaoar and Sedna. We detail the evidence for the presence of such ices below and summarize it in Table 1. See deBergh et al 2013, Brown 2008, 2012, and Grasset et al. 2017 for earlier reviews, and the chapter by Barucci and Merlin (2019, this volume) for more general discussion of the composition of TNO surfaces. Despite attempts to characterize surface compositions with photometric systems (Trujillo et al. 2011, DalleOre et al. 2015), spectral resolution of at least 500 seems to be needed for definitive detections of volatile ices, and we focus on spectral data below.

*Triton:* We consider Triton in the discussion of volatiles on Trans Neptunian Objects, even though it is a moon of Neptune, because it is thought to be a captured TNO (McKinnon & Leith 1995). and, more frivolously, its orbit clearly crosses that of Neptune. Its surface and atmospheric properties are usefully compared to those of Pluto and other large TNOs. The volatiles, $N_2$ and $CH_4$, were measured in Triton's atmosphere by Voyager during its 1989 flyby with its UV spectrometer (Broadfoot et al. 1989). Three relatively volatile ices, $N_2$, $CH_4$, and CO, and two non-volatile ices, $CO_2$ and $H_2O$, were detected on Triton's surface from ground-based near-IR spectroscopy (Cruikshank et al. 1993, Cruikshank et al. 2000). Noting that the spectral profile of the 2.15-$\mu$m $N_2$ absorption band is temperature dependent, and transitions from very broad for the higher-temperature hexagonal β-$N_2$ phase above 35.61 K to very narrow for the colder cubic α-$N_2$ phase, Quirico et al. (1999) showed that the $N_2$ ice was in the warmer β phase on Triton. Additionally, the spectrum of $CH_4$ ice diluted in $N_2$ is shifted toward shorter wavelengths with respect to pure $CH_4$ ice (Quirico & Schmitt 1997; Protopapa et al. 2015). Therefore it was concluded that solid $CH_4$ on Triton exists predominantly diluted in $N_2$ (Quirico et al. 1999), with subsequent observations suggesting the presence of some pure $CH_4$ ice (Merlin et al. 2018). These ices are not uniformly distributed across Triton's surface, as seen from the variation of its spectrum with sub-observer longitude as the body rotates (Grundy & Young 2004, Grundy et al. 2010, Holler et al. 2016). These studies suggest that the non-volatiles $H_2O$ and $CO_2$ may dominate the terrains nearest Triton's current summer pole, that the more volatile species, $N_2$ and CO, are co-located, and that the $CH_4$ distribution and mixing state may vary with depth and longitude. Furthermore, there is evidence for isolated areas of $CH_4$ diluted in $N_2$ that is too fine-grained to allow direct detection of the inherently weak 2.15 $\mu$m $N_2$ band, where the presence of the $N_2$ is revealed only by its effect of shifting the $CH_4$ spectrum. Further clues to the distribution of Triton's ices comes from a long time base of observations and Triton's changing aspect. Infrared spectra from 1986 to 1992 — approaching Triton's summer solstice in 2000 — show a dramatic decrease in Triton's 2.2-2.4 $\mu$m $CH_4$ band (Brown et al. 1995), while post-solstice spectra 2002-2014 show a mild increase in $CH_4$ band depth (Holler et al. 2016), suggesting a combination of changing aspect and perhaps active volatile transport. Voyager flew no infrared spectrometer, and so it is likely that these speculations will remain unconfirmed until a mission to Neptune/Triton can fly an infrared spectrometer.

*Pluto:* Pluto's surface has been studied with disk-integrated ground-based spectra in the visible, near infrared and mid infrared (see the review by Cruikshank et al. 2015) and at geologically relevant spatial scales by the LEISA infrared spectrometer and MVIC color imager on NASA's New Horizons spacecraft (see chapter by Spencer et al.). The $N_2$ ice on Pluto is concentrated in a large, deep, $N_2$-filled basin called Sputnik Planitia, near the equator and opposite Charon, in which some CO and $CH_4$ is diluted in solid solution with large-grained or annealed $\beta$-$N_2$. $N_2$-rich ice of similar composition is also seen at mid latitudes, 35-55° N (Protopapa et al. 2017; Schmitt et al. 2017). It is now understood that Pluto has more $CH_4$ than Triton overall. Areas of $CH_4$-rich ice was first inferred from ground-based spectra, and were mapped with the LEISA spectrometer to be predominately (i) at a northern cap, north of ~55°N, (ii) in a band from 20-35°N, and (iii) on the area named Tartarus Dorsae on the eastern terminator limb of the encounter hemisphere near 220° to 250° E, 10°S to 30° N.

*Eris:* The visible and near-infrared spectrum of Eris clearly shows strong $CH_4$ absorption features (Brown et al 2005; see review by Brown 2008 and analysis by Tegler et al. 2012). Absorption at 1.684 $\mu$m indicates pure $CH_4$ or $CH_4$-rich ice (Dumas et al. 2007), and the spectral shifts are much smaller than for either Triton or Pluto spectra. The depth, shape, and the precise wavelengths of the $CH_4$ features on Eris can be modeled to derive grain size, dilution state, and stratification of $CH_4$ ice, even without the direct detection of the $N_2$ feature (e.g., Merlin et al. 2009). The issue of stratification—i.e., whether the $CH_4$ concentration state is constant or varying with depth—is particularly interesting because it relates to the evolution of Eris's ices over its 561-year orbit, which in turn is determined by the order in which different species freeze out onto the surface each orbit after perihelion. Different studies have disagreed on the stratification, variously concluding that the dilute $CH_4$ lies above pure $CH_4$ (Licandro et al. 2006a), below pure $CH_4$ (Abernathy et al. 2009), sandwiched between two layers of pure $CH_4$ (Merlin et al. 2009), or that the stoichiometry of $CH_4$:$N_2$ is constant with depth (Tegler et al. 2012). Clearly, better data and continued modeling is needed on this question. Eris's surface temperature is almost certainly below the $\alpha$-$\beta$ phase transition for $N_2$ (Sicardy et al. 2011), given its bright geometric albedo $p_V = 0.96$. Since the $\alpha$-$N_2$ absorption feature is weaker and much narrower than that of $\beta$-$N_2$, it has eluded direct detection on Eris.

*Makemake:* Very strong $CH_4$ absorption is seen in the visible and infrared spectrum of Makemake with spectral shifts to shorter wavelengths, but with smaller shifts than are seen on Eris, Pluto, or Triton (Licandro et al. 2006b, Brown et al. 2007, Tegler et al 2007, 2008, Lorenzi et al. 2015, Perna 2017). This suggests that although some $CH_4$ is in solution with $N_2$, $N_2$ itself is not as prevalent on Makemake as on the previous three bodies. At visible wavelengths, Tegler et al (2007) detected four absorption features between 0.54 and 0.62 $\mu$m that they attributed to $CH_4$, but ice-phase laboratory spectra of these features are lacking. The non-volatile irradiation products of $CH_4$ are also seen on Makemake (Brown et al. 2015).

*Quaoar:* The spectrum of Quaoar is dominated by $H_2O$ ice, analogous to $H_2O$-rich Charon. Jewitt & Luu (2004) interpreted absorption near 1.65 $\mu$m as crystalline water ice and absorption near 2.2 $\mu$m as ammonia hydrate, and reported no detected $CH_4$ absorption. Improved SNR at 2.2-2.4 $\mu$m strengthened the case for $CH_4$ in the spectrum (Schaller & Brown 2007b), which, if real, was nevertheless subtle when superimposed on the significant $H_2O$ absorption. Guilbert et al. (2009) claim a marginal detection (2 ± 2% deep) of the 1.724 $\mu$m feature due to $CH_4$ ice, which they say adds weight to the attribution of $CH_4$ ice for the 2.2 $\mu$m feature. Dalle Ore et al. (2009) model the near-IR spectrum plus photometry at 3.6 and 4.5 $\mu$m using $CH_4$ to explain both the 2.2 $\mu$m absorption and the mid-IR photometry, with no need for ammonia hydrate, and propose that $N_2$ ice in the β-phase may cover 20% of the surface, as seen by relatively subtle effects near 2.15 $\mu$m and by photometry at 3.6 and 4.5 $\mu$m. Barucci et al. (2015) investigate the 1.67 $\mu$m $CH_4$ band, and $N_2$ implied by the shifts in the 1.67 $\mu$m $CH_4$ band, but measuring this shift is complicated by the broad 1.65 $\mu$m crystalline $H_2O$ band. In short, on Quaoar strong $H_2O$ bands complicate the definitive detection and analysis of $CH_4$, and the derived presence of $N_2$ from $CH_4$ shifts.

*2007 $OR_{10}$*: The large TNO 2007 $OR_{10}$ (Pál et al. 2016) shows an extremely red visible spectrum, and its near-IR spectrum shows water-ice absorption (Brown et al. 2011). 2007 $OR_{10}$ shares both traits with Quaoar. This similarity with Quaoar is indirect evidence that $CH_4$ may also be present on 2007 $OR_{10}$ (Brown et al., 2011), if one accepts both the equivalence of surface type, and the evidence of $CH_4$ on Quaoar. Both Quaoar and 2007 $OR_{10}$ are near the $CH_4$ retention line of Schaller and Brown (2007a), see Section 4.

*Sedna*: Distant Sedna, currently at ~85 AU, has a spectrum that is difficult to interpret. Barucci et al. (2005) reported hints of a Triton-like spectrum with β-$N_2$ (an unexpected $N_2$ phase at Sedna's distance) and $CH_4$, from spectra with a spectral resolution of only 100 in the near-IR, while Trujilo et al. (2005) reported a featureless spectrum, from spectra at a binned resolution of 215 near 2.15 $\mu$m. Near-infrared spectra were analyzed in combination with Spitzer photometry at 3.6 and 4.5 μm (Barucci et al. 2010, Emery et al. 2007). No diagnostic $CH_4$ absorption bands were evident, but a decrease in albedo longward of 2.2 $\mu$m was interpreted as due to $CH_4$ and $C_2H_6$ (ethane), or possibly serpentine, and the addition of $N_2$ was also consistent with the Spitzer photometry (Barucci et al., 2010). The surface appears heterogeneous (Barucci et al. 2010).

*Varuna*: The spectrum of Varuna is in general well explained by various non-volatiles, such as water ice, olivine, pyroxene, tholin, and amorphous carbon as a darkening agent. The strong $CH_4$ absorption bands at 1.67, 1.72, 1.8, or 2.21 $\mu$m are not detected in the current best spectra (Lorenzi et al. 2014), limiting the amount of $CH_4$ ice to less than 10% of the surface. Lorenzi et al. (2014) discuss how the addition a small amount methane ice slightly improves the match to the data, but as their data have a spectral resolution of only 50, further work is critically needed to confirm or constrain $CH_4$ on Varuna.

*Orcus*: The water-dominated near-infrared spectrum of Orcus (de Bergh et al. 2005, Fornasier et al. 2004, Trujillo et al. 2005) has an absorption feature near 2.2 $\mu$m, compatible with absorption by $CH_4$, $NH_3$, or $NH_4^+$ (Barucci et al. 2008; Delsanti et al. 2010; Carry et al. 2011; DeMeo et al. 2010). Because of the lack of other $CH_4$ absorptions, specifically those near 1.67 and 1.72 μm, it is likely that $CH_4$ is not the main cause of the 2.2 $\mu$m absorption feature.

# INSERT TABLE 1 HERE

## 3. Volatile-supported atmospheres

For a TNO with little or no atmosphere, the equilibrium surface temperature, $T_0$ will depend on its rotation rate and thermal inertia (i.e. fast vs. slow rotator), latitude ($\lambda$), sub-solar latitude ($\lambda_{sol}$), and time of day (e.g., hour angle of the sun, $h$). The simplest approach is to assume thermal emission balances absorbed insolation, and that the atmosphere is nearly transparent, in which case the equilibrium surface temperature is:

$$(1) \qquad \epsilon \sigma T_0^4 = \overline{\mu(\lambda, \lambda_{sol}, h)} \left( \frac{S_{1AU}(1-A)}{\eta R^2} \right)$$

where $\epsilon$ is the emissivity, σ is the Stefan–Boltzmann constant, $S_{1AU}$ is the normal solar insolation at 1 AU (1367 W m$^{-2}$), $A$ is the bolometric Bond albedo, and $R$ is the heliocentric distance in AU. $\overline{\mu(\lambda, \lambda_{sol}, h)}$ is the average cosine of the incidence angle for the problem at hand: for the sub-solar temperature on a slow rotator, $T_{ss}$, $\bar{\mu} = 1$; for the equatorial latitude for fast rotator, $\bar{\mu} = 1/\pi$; for a single characteristic equilibrium temperature balancing the global average of the insolation, $T_{eq}$, $\bar{\mu} = 1/4$. $\eta$ is the so-called "beaming factor" (e.g., Spencer 1990, Harris 1998, Müller et al. 2010, Lellouch et al. 2013), which can raise the temperature for a rough surface (or lower it for a body with high thermal inertia in some models of thermal emission). Many of the smaller TNOs have $A \approx 5\%$, while the largest TNOs have $A \approx 60\text{-}80\%$ (Lellouch et al. 2013). Thus, the temperatures relevant for TNOs range from $T_{ss} \approx 70$ K for $\epsilon=1$, $\eta=1$, $A = 5\%$ at $R=30$ AU down to $T_{eq} \approx 20$ K for $\epsilon = 1$, $\eta = 1$, $A = 80\%$ at 90 AU. Detached objects at $R$ much greater 90 AU will have even colder temperatures. Eq (1) ignores latent heat of sublimation and internal heat flux, and the conduction of heat into or from the subsurface is empirically included in the beaming factor.

The surface pressure of an atmosphere over a pure ice in thermodynamic equilibrium (the sublimation pressure, $p_s(T_0)$) is a function only of the ice temperature (Fray & Schmitt 2009). Of the three super-volatiles seen in the outer solar system, $N_2$, CO, and $CH_4$, $N_2$ is by far the most volatile (Fig 1). TNOs with volatiles show a mix of ices: $CH_4$ or CO diluted in $N_2$-rich ice; $N_2$ or CO diluted in $CH_4$-rich ice; or pure $CH_4$ ices. The mixtures present complicated surface-ice interaction, including lag deposits and the influence of $CH_4$-rich warm patches (see review by Trafton et al. 1998). Trafton (2015) and Tan and Kargel (2018) have more recent work on the mixtures, including the exciting conclusion that the $CH_4$ and $N_2$ partial pressures above a mixture of $CH_4$ saturated in $N_2$-rich ice plus $N_2$ saturated in $CH_4$ ice is independent of the bulk $N_2$:$CH_4$ ice ratio, but these have not been tested under laboratory conditions. Moreover, as seen in the compilation by Fray and Schmitt (2009), laboratory measurements for the pure ices only exist for temperatures above 54.78 K for CO and above 48.15 K for $CH_4$; and the only laboratory data below 35.4 K for $N_2$ were published in 1960 (Borovik et al. 1960).

## INSERT FIG 1 HERE

The surface pressure is the weight of the column of gas, and so is closely related to the column density at the surface, $N_0$

(2) $$p_s(T_0) \approx mg_0N_0 \approx mg_0H_0n_0$$

where $m$ is the mass of molecule, and $g_0$ is surface gravity, given by $g_0 = GM/r_0^2$, where $G$ is the gravitational constant, $M = \rho(4/3)\pi r_0^3$ is the TNO mass, $\rho$ is the TNO bulk density, and $r_0$ is the surface radius. $H_0 = kT_0/mg_0$ is the pressure scale height at the surface, $k$ is Boltzman's constant, and $n_0$ is the local number density at the surface. The relation is only approximate for small bodies, because gravity decreases with altitude.

For small column densities, the atmosphere can be described as a surface-bounded exobase, since if $N_0 \sigma_{eff} \ll 1$, where $\sigma_{eff}$ is the effective collision cross section, then an escaping molecule is not likely to suffer a collision on its way out. This condition is equivalent to a large Knudsen number, $Kn$, the ratio of the mean free path between collisions, $l_{mfp} = 1/(n_0 \sigma_{eff})$, to the a characteristic length scale. The scale in question depends on the problem at hand; taking the scale height for the length scale (Zhu et al. 2014), $Kn = 1$ is one classic definition of an exobase. At the surface:

(3) $$Kn_0 = \frac{l_{mfp}}{H_0} = \frac{1}{N_0\sigma_{eff}}$$

$\sigma_{eff}$ is $\sqrt{2}$ times larger than the collisional cross section, $\sigma_{coll}$ (Johnson et al. 2015). Since $\sigma_{coll} \approx 0.46 \times 10^{-14}$ for $CH_4$ (Atkins and de Paula 2009) or $\sigma_{coll} \approx 0.43 \times 10^{-14}$ for $N_2$ (Kaye and Laby 1973), with some dependence on temperature, atmospheres with $N_0$ greater than $\sim 1.6 \times 10^{14}$ molecule $cm^{-2}$ can be considered collisional. This is achieved for $N_2$ or $CH_4$ at extremely small surface pressures (Fig 1), e.g., $\sim 3 \times 10^{-8}$ $\mu$bar for $CH_4$ on a $\rho=1.6$ g $cm^{-3}$, 100 km radius body, or $\sim 8 \times 10^{-7}$ $\mu$bar for $N_2$ on a $\rho = 2.5$ g $cm^{-3}$, 1000 km radius body. Extrapolating to these small pressures from the pressures measured in the lab is questionable, but application of the Fray and Schmidt compilations predicts that the transition to continuum atmospheres happens near temperatures of $\sim 29$ K for $CH_4$, and $\sim 21$ K for $N_2$, depending weakly on the TNO radius and density. For larger column densities, the atmosphere becomes opaque to UV radiation. This transition occurs for Ly-$\alpha$ at $N_0 \sim 10^{18}$ molecule $cm^{-2}$ ($T_0 \sim 26$ K) for $N_2$ with 3% $CH_4$ gaseous molar mixing ratio, or $N_0 \sim 3 \times 10^{16}$ molecule $cm^{-2}$ ($T_0 \sim 32$ K) for pure $CH_4$ (Johnson et al. 2015).

The Jeans parameter at the surface, or ratio of potential energy to thermal energy, is a measure of how tightly bound the atmosphere is, and is given by:

(4) $$\lambda = \frac{gm}{rkT} = \frac{U}{kT} = \frac{r}{H} = \frac{GMm}{r^2}$$

The Jeans parameter at the surface, $\lambda_0$, varies from $\sim 0.1$ (unbound) for $CH_4$ gas on 100-km radius objects at 70 K to $\sim 100$ (bound) for $N_2$ gas on 1000-km radius objects at 20 K. Although the composition surely varies from object to object, assuming each TNO has $N_2$ on its surface it can be seen from Fig. 2 that due to the large range of temperatures, the TNO atmospheres can have a large range of $\lambda_0$ and $N_0$, with implications for atmospheric escape, seasonal variation, and detectability.

## INSERT FIG 2 HERE

## 4. Expected volatile retention

In general, only the largest TNOs have had volatiles detected or suspected on their surfaces (Table 1). This is not merely an observational effect (i.e., because higher quality spectra are more easily obtained on larger TNOs), but is linked to the escape of their initial inventory of volatiles (Schaller & Brown 2007a; Levi and Podolak 2009; Johnson et al. 2015). Volatile escape can be driven by heating of the surface by solar visible radiation and by absorption of solar radiation in the atmosphere. The relative importance of these processes depends on how tightly bound the atmosphere is, parameterized by the surface Jeans parameter $\lambda_0$, and its UV opacity, parameterized by its surface column density, $N_0$. Since the work of Schaller & Brown (2007a) and Levi and Podolak (2009) there has been progress in estimating the escape rate in the transition to the fluid regime for transparent atmospheres (Volkov et al. 2011a,b), and the role of atmospheric heating on escape (Johnson et al 2013a,b, 2015), as well as new observations relating to initial volatile inventories (Glein & Waite 2018) and the complexity of escape at Pluto (Gladstone & Young 2019).

In order to provide a link between the presence of volatiles, the bulk properties, and the orbits of TNOs, Schaller & Brown (2007a) considered sublimation-induced escape directly from the TNO surface, and updated this work in Brown et al. (2011). Starting with the equilibrium surface temperature (e.g., Eq. 1 with $\bar{\mu} = 1/4$ and $\eta = 1$), they used the Jeans expression for escape from an exobase, evaluated at the conditions of the surface (subscripted here as *SJ* for surface-Jeans):

(5) $$\Phi_{SJ} = 4\pi r_0^2 n_0 (\bar{v}/4)(1 + \lambda_0)\exp(-\lambda_0)$$

where $\Phi$ is the total escape rate of volatiles in molecule s$^{-1}$, and $\bar{v} = \sqrt{8kT/\pi m}$ is the mean molecular speed. Using these expression for $T_0$ and $\Phi$, they divided the TNOs into those that would likely keep their volatiles and those that likely lost their initial volatile inventory over the age of the solar system (Fig 3).

## INSERT FIG 3 HERE

The surface-Jeans estimate of the escape rate is roughly accurate if the atmosphere is non-collisional, $N_0 < \sim 10^{14}$ molecule cm$^{-2}$ (Fig 4), which holds for very distant TNOs. However, Eq. 5 is problematic, even for atmospheres that are transparent to solar heating. Levi and Podolak (2009) subsequently used a hydrodynamics model that transitioned to Jeans escape. Volkov et al. (2011a, b) used a molecular kinetic model, the Direct Simulation Monte Carlo (DSMC) method (Bird 1994), to calculate the surface-heated escape rate, $\Phi_S$, from a single-component atmosphere for a range of surface values of $T_0$ and $N_0$, and then expanded this range to very thick atmospheres by coupling iteratively to a fluid model in (Johnson et al. 2015), thereby covering the full range of escape due to surface heating. Note that Volkov et al. (2011a, b) use the *radial* Knudsen number, scaling the mean-free-path by the surface radius, which is appropriate for small bodies with extended atmospheres (i.e., small $\lambda_0$). We denote that here as $Kn_0^r$ (*r* for radial) and relate it to Eq (3) through $Kn_0^r = l_{mfp}/r_0 = Kn_0 / \lambda_0$. Johnson et al. (2015) fit an analytic expression to the numerical results of Volkov et al. (2011a, b) to find a correction to the surface-Jeans flux that depends on the surface values of the Knudsen number and Jeans parameter. We use $Kn_0^r$ here to allow direct comparison with Volkov et al. (2011a, b) and Johnson et al. (2015).

(6) $$\Phi_S = \Phi_{SJ}/[(Kn_0^r)^{0.09} + \lambda_0^{2.55}\exp(-\lambda_0)/(70Kn_0^r)]$$

For cold atmospheres and large bodies (high gravity), the escape rate driven only by the surface temperature is throttled for a bound atmosphere by the exp(-$\lambda_0$) term in Eq. (5), (Fig 4). This can be overcome by direct absorption of solar radiation in the atmosphere. Assuming ~2-3% $CH_4$ in the more volatile $N_2$ background gas, Johnson et al. (2015) calculated that the column of gas that is sufficient for the UV and EUV to be primarily absorbed in the atmosphere is $N_0 > N_C$, where $N_C \approx 10^{18}$ molecule cm$^{-2}$ is the minimum column for UV absorption. Prior to the New Horizons encounter, models of Pluto's atmospheric loss coupled a fluid simulation to a DSMC molecular kinetic simulation (e.g., Erwin et al. 2013; Tucker et al. 2012) to describe the UV/EUV absorption vs. depth as well as the escape from the exobase region. In their Pluto modeling, the simulated loss rate was matched reasonably well by the simple so-called energy limited escape model, in which the gravitational energy lost by the escaping molecules is set equal to the heating produced by the absorbed UV radiation. The limits to the applicability of that model was explored in Johnson et al. (2013) and subsequently applied to other KBOs (Johnson et al. 2015). Ly$\alpha$ dominates the absorbed UV flux if $CH_4$ is optically thick. Since scattered interplanetary Ly$\alpha$ and stellar flux contribute to the UV flux in the trans-Neptunian region, the energy-limited flux falls off more slowly than $R^2$. Eq. (7) gives the expression adopted by Johnson et al. (2015) for the escape due to UV/EUV heating in the upper atmosphere, $\Phi_U$, scaled to the detailed rate calculated for Pluto, $\Phi_P$:

(7) $$\Phi_U = \Phi_P(\rho_P/\rho)[(30/R)^2 + 0.09]$$

where $\rho_P$ and $\rho$ are is the density of Pluto and the TNO, $R$ is the heliocentric distance in AU, and $\Phi_P$ is the escape rate at Pluto due to the direct solar UV flux. The second term in the brackets accounts for so-called interplanetary UV flux (Gladstone et al. 1998) ignored in those papers, but which becomes important at very large $R$. Johnson et al. (2015), based on Erwin et al. 2013 and Tucker et al. 2012, adopted $\rho_P = 2.05$ g cm$^{-3}$, and $\Phi_P = 120$ kg s$^{-1}$ / $m_{N2} = 2.6 \times 10^{27}$ $N_2$ s$^{-1}$. Johnson et al. (2015) combined surface and upper atmospheric heating by restricting $\Phi_U$ to $N_0 > N_C$ and choosing the larger of $\Phi_S$ and $\Phi_U$ (Fig 4). Fig. 4 only gives a rough guide for ranges in which global models of the two escape processes dominate. Johnson et al. (2015) integrated the combined atmospheric loss rates over the lifetime of individual TNOs (Table 2 in Johnson et al. 2015, which did not include Orcus and Varuna). Besides Charon, which has likely lost its initial volatiles, they concluded that for the objects studied only Makemake, Quaoar and 2007 OR$_{10}$ likely lost a large fraction of their volatiles, primarily due to short wavelength absorption in their upper atmospheres, while Pluto, Triton, and Sedna had retained most of theirs. While similar conclusions were reached by Schaller & Brown (2007a), the fraction of volatiles retained by Pluto, Triton, and Sedna differed.

## INSERT FIG. 4 HERE

The picture created by these simulations has been altered by two recent spacecraft measurements. One is our more recent understanding of Pluto's atmosphere based on the New Horizons observations (Gladstone and Young 2019). Prior to the flyby, the expected escape rate was ~[0.4-4] x $10^{27}$ $N_2$ $s^{-1}$ (Zhu et al. 2014), consistent with energy-limited escape. The escape rate that is derived from the observed density and temperature profile is much lower: (3–7) × $10^{22}$ $N_2$ $s^{-1}$ and (4–8) × $10^{25}$ $CH_4$ $s^{-1}$ (Young et al 2018). Applying these new observations has been confounded by the fact that the principal cooling agent in Pluto's upper atmosphere, which also compressed the extent of Pluto's atmosphere, is still uncertain (Gladstone & Young 2019). Pluto's current escape rate must be better understood before volatile loss can be calculated for Pluto at other epochs, or for other bodies. The second is new constraints on the initial inventory of volatiles. While Schaller & Brown 2007a and Johnson et al. 2015 used an $N_2$ to $H_2O$ mass ratio of 2%, Glein & Waite (2018) use the mixing ratio of $N_2$ measured in the coma of 67P (Rubin et al., 2015) to derive an $N_2$ to $H_2O$ mass ratio of only [0.7-6]x$10^{-4}$.

## 5. Variation of atmospheres over an orbit

Because the sublimation pressures depend exponentially on the temperatures of the volatile ices, the gases surrounding volatile-bearing TNOs vary with heliocentric distance and subsolar latitude, and possibly time of day and latitude. This was initially modeled for Triton and Pluto (see reviews by Spencer et al. 1997 and Yelle et al. 1995). Since those reviews, trends of increasing atmospheric pressure for both Triton and Pluto were observed using the technique of stellar occultation, with an increase by factors of two and three respectively (Elliot et al. 1998; Elliot et al. 2000; Elliot et al. 2003a; Olkin et al. 1997, 2015; Meza et al 2019; see section 6). The new time-base of atmospheric observations and the New Horizons flyby of Pluto inspired new models of seasonal variation (e.g., Young 2012, 2013, 2017; Hansen et al. 2015; Olkin et al. 2015), including general circulation models (e.g., Forget et al. 2017) and evolution of atmospheres on the timescale of millions of years (e.g., Bertrand et al. 2016, 2018).

When $N_2$ was discovered on the surface of Eris, authors speculated that volatiles on TNOs, especially $N_2$, could raise temporary atmospheres near perihelion (e.g., Dumas et al. 2007). This was generalized in Stern and Trafton (2008), and applied numerically to the known or suspected volatile-bearing TNOs by Young & McKinnon (2013). When thinking about atmospheres on TNOs, it is useful to distinguish three types: global, collisional, and ballistic. For global sublimation-supported atmospheres, such as Mars or current-day Pluto and Triton, volatiles sublime from areas of higher insolation, and recondense on areas of lower insolation, transporting latent heat as well as mass (Trafton 1984, Ingersoll 1990; also see reviews by Spencer et al. 1997; Yelle et al. 1995; Stern and Trafton 2008). As long as the volatiles can be effectively transported, the surface pressures and the volatile ice temperatures will be nearly constant across the surface. Sublimation winds transport mass from latitudes of high insolation to low insolation. Trafton (1984) showed that pressures stay within 10% across the surface if the sublimation winds ($v$) are less than 7.2% of the sound speed ($v_s$). The sublimation wind speeds can by found by conservation of mass: the mass per time crossing a given latitude equals the integral of the net deposition from that latitude to the pole. The wind speeds depend on the subsolar latitude (Trafton 1984), if we consider diurnally averaged insolation; higher wind speeds are needed to transport volatiles pole-to-pole (high subsolar latitudes) than equator-to-pole (low subsolar latitudes). For an "ice ball" uniformly covered in volatiles, the maximum sublimation wind speed, $v$, can be expressed as

(8) $$vmN_0 = \xi Sr/L$$

where $S = S_{1AU}(1-A)/R^2 = 4\varepsilon\sigma T_{avg}^4$ is the absorbed normal insolation, and $L$ is the latent heat of sublimation, in energy per mass. $\xi$ in Eq. 8 is a numerical factor accounting for the subsolar latitude, $\lambda_{sun}$. We calculated $\xi$ numerically, following the prescription of Young (1992). From these calculations, $\xi$ is well approximated by a cubic expression

(9)  $\xi(\lambda_{sun}) \approx 0.044 + 0.148\,(\lambda_{sun}/90°) + 0.4012\,(\lambda_{sun}/90°)^2 - 0.296\,(\lambda_{sun}/90°)^3$

For a 400-1400 km radius body uniformly covered with $CH_4$ ice to have a global atmosphere, the pressure needs to be greater than ~17 to 295 nbar for polar illumination ($3.1\times10^{19}$ to $1.6\times10^{20}$ cm$^{-2}$, 41.0 to 45.6 K; Fig 5), or 1.9 to 33 nbar for equatorial illumination ($3.6\times10^{18}$ to $1.8\times10^{19}$ cm$^{-2}$, 38.1 to 42.0 K). For $N_2$, the pressures are similar, so the temperatures are lower: 14 to 244 nbar for polar ($1.5\times10^{19}$ to $7.4\times10^{19}$ cm$^{-2}$, 29.1 to 31.9 K) or 2 to 28 nbar for equatorial ($1.8\times10^{18}$ to $68.5\times10^{18}$ cm$^{-2}$, 27.2 to 29.7 K). $N_0$ increases slightly faster than $r$ because both $S$ and $N_0$ increase with temperature; $p_0$ increases even faster, slightly faster than $r^2$, because of its dependence on $g_0$ (Eq. 2).

The temperatures in Fig 5 are highly simplified. Seasonal thermal inertia can be important, even at the long timescales in the outer solar system. More significantly, bodies are unlikely to be uniformly covered in volatiles. For example, much of the $N_2$ on Pluto is located in the basin known as Sputnik Planitia (Moore et al. 2016), and Triton's $N_2$ may be perennially confined to the southern hemisphere (Moore & Spencer 1990).

## INSERT FIG. 5 HERE

Non-global atmospheres will vary with location and time-of-day, but may still be collisional, if the column density is greater than ~$10^{14}$ cm$^{-2}$ for either $N_2$ or $CH_4$. Io is a classic example of a local atmosphere that is collisional around the sub-solar point, and demonstrates some of the processes that are active in even these thin atmospheres. Atmospheric chemistry can occur even in these local, tenuous atmospheres (Wong & Smyth 2000). Supersonic winds certainly flow and transport volatiles, even if they are not effective at equalizing pressures and temperatures (e.g., Walker et al. 2012). Recently, Hofgartner et al. (2018) used the Ingersoll et al. (1985) meteorological model developed for Io study the transport of $N_2$ on Eris at aphelion, when it is a local, collisional atmosphere, and found significant transport of $N_2$ ice. Even for more tenuous "surface-bounded exospheres," the loss of volatiles can modify landforms (see review by Mangold 2011). For example, sublimation erosion may lead to the narrow divides between craters on Hyperion (Howard et al. 2012) or redeposition on the crater rims on Callisto, where the convex summits see less of the warm surface than do concave crater interiors, and are therefore local cold traps (Howard and Moore 2008).

### 6. Detections of or limits on atmospheres by stellar occultation.

The technique of stellar occultation is one of the most powerful ways to search for atmospheres around these bodies. While infrared absorption or radio emission have detected low column densities of CO and $CH_4$ on Triton (Lellouch et al. 2010) or CO and HCN on Pluto (Lellouch et al. 2017), these are bright targets. Stellar occultations depend on the brightness of the occulted star, and also study the target's size, shape, and the presence or nature of rings or jets (see Ortiz et al. 2019, this volume; Elliot and Olkin 1996, Elliot & Kern 2003; Santos-Sanz et al 2016). In an occultation, the starlight is refracted through a bending angle that increases in magnitude roughly in proportion to the line-of-sight column density ($N_{los}$). This leads to differential refraction, or a divergence of the refracted rays (cf Elliot and Olkin 1996), which dims the occulted starlight according to the scale height $H$ and the column density. Some insight can be gained (Fig. 6) from the approximate, analytic expression for the relative stellar flux, $\phi$, (e.g., Elliot and Young 1992):

(10) $$\phi \approx \left|1 - \theta \frac{\Delta}{H}\right|^{-1} \left|1 + \theta \frac{\Delta}{r}\right|^{-1} e^{-\sigma_{ext} N_{los}}$$

where $\theta$ is the bending angle, given by $\theta \approx -\nu_{STP} N_{los}/n_{STP} H$, $\nu_{STP}$ is the refractivity at standard temperature and pressure ($2.9 \times 10^{-4}$ and $4.4 \times 10^{-4}$ for $N_2$ and $CH_4$ at 0.7 $\mu$m), $n_{STP}$ is Loschmidt's constant ($2.6868 \times 10^{19}$ cm$^{-3}$), and $\Delta$ is the target-observer distance. As before, $r$ is the radius (distance from target center) and $H$ is the atmospheric scale height. $N_{los}$ is the line-of-sight column density (molecule per area), which is larger than $N$, the vertical column density, by the unitless factor $\sqrt{2\pi\lambda}$ (e.g. 8-25 for $\lambda \approx$ 10-100). $\sigma_{ext}$ is the extinction cross section (area per molecule). The first term represents the decrease due to the divergence (defocusing) of rays perpendicular to the limb, and halves the starlight when $\theta\Delta \approx -H$. The second term represents the increase due to refocusing parallel to the limb, leading to a "central flash" near the center of the shadow. The final term represents extinction, which becomes important when $N_{los} \approx 1/\sigma_{ext}$ (e.g., $N_{los} \approx 10^{26}$ molecule cm$^{-2}$ for Rayleigh scattering at visible wavelengths, or $N_{los} \approx 10^{17}$ molecule cm$^{-2}$ for typical crosssections in the extreme ultraviolet (UV)). Refraction is typically dominant over extinction for Earth-based stellar occultations of TNOs, and extinction dominates for spacecraft UV occultations.

## INSERT FIG. 6 HERE

Lightcurves from model atmospheres can be calculated analytically for some idealized atmospheres (e.g., isothermal or $T \propto r^\beta, \beta \ll 1$, Elliot and Young 1992). However, the presence of $CH_4$ or its by-products (including photochemical haze) can heat up the atmosphere by 10s of K (Yelle and Lunine 1989, Zhang et al. 2018). For more complex atmospheres, synthetic lightcurves are calculated under the assumption of geometric optics/ray-tracing (Sicardy et al. 1999 and references therein) or wave optics/Fresnel diffraction (French and Gierasch 1976). Standard model fitting can then be used to extract the geometric edge and the refractivity of the atmosphere at the surface. For high quality data (as has been obtained for Pluto and Triton), lightcurves can be inverted to extract temperature, pressure, and number-density profiles (e.g. Elliot et al. 2003b). Derived surface pressures (or upper limits) from occultation lightcurves can be compared to the sublimation pressures for the atmospheric molecules under consideration, as expected from its surface equilibrium temperature (Fray and Schmitt 2009).

Synthetic light curves vs. shadow radius are plotted in Fig. 7 for an example TNO with an $N_2$ or $CH_4$-dominated atmosphere (see caption for model details). Very thin atmospheres (10's to 100's of nanobar, or ~$10^{19}$ to $10^{20}$ molecule cm$^{-2}$) cause only a small drop of flux very close to the surface, requiring very high signal to noise ratio for detection. Denser atmospheres (a few $\mu$bar, or ~$10^{20}$ molecule cm$^{-2}$) cause a gradual drop in the star flux at a significant distance from the object's surface, so they will be easier to detect. For denser atmospheres, the ray that grazes the TNO surface is refracted inward significantly towards smaller shadow radii (bent by 94 to 112 km for 1 $\mu$bar $CH_4$ or $N_2$ atmospheres). For the 10 $\mu$bar curve in Fig. 7, the surface-grazing ray is refracted past the shadow center. The bottom flux never reaches zero and much of the lightcurve is the sum of the near and far limb contributions. In the example plotted, the limb-grazing ray is bent by 900 to 1180 km for a $CH_4$ or $N_2$ atmospheres, leading to the discontinuities seen at ±400 km or ±680 km for the $CH_4$-or the $N_2$-dominated atmospheres. When the star crosses the center of the object as seen from Earth, it causes a prominent central flash for the 10 microbar atmosphere, as the atmosphere acts as a lens, converging the starlight from all parts of the atmosphere to the observer.

## INSERT FIG 7 HERE

Stellar occultations by Pluto, Charon, and Triton, have been observed since the 1980's.

*Pluto*: A single-chord occultation from 1985 was observed under extremely difficult circumstances (Brosch 1995). The definitive discovery of Pluto's atmosphere was from the 1988 stellar occultation (Hubbard et al. 1988, Millis et al. 1993). Observations in 2002 showed a doubling of the pressure since 1988 (Sicardy et al 2003; Elliot et al. 2003a). Many high-quality observations since have revealed the continued changes in Pluto's atmosphere, its thermal structure, and waves (Young et al. 2008; Toigo et al. 2010; Sicardy et al. 2016; Dias-Oliveira et al. 2015; Pasachoff et al. 2017; Meza et al. 2019). The New Horizons radio and solar occultations revealed $N_2$ in vapor-pressure equilibrium with the $N_2$-rich ice, an overabundance of gaseous $CH_4$ possibly explained by $CH_4$-rich patches, hazes and photochemical products, a cold upper atmosphere, and an atmospheric escape rate dominated by $CH_4$ and much smaller than expected (See Spencer et al. 2019, this book, and Gladstone & Young 2019).

*Triton*: Triton, like Pluto, was visited by a spacecraft, and has had several high-quality occultations, notably the 1997-Nov-4 Triton occultation observed from HST (Elliot et al. 1998). Occultations in the 1990's show a doubling of surface pressure on decadal timescales compared to the 1989 flyby by Voyager 2 (Elliot et al. 1998; Elliot et al. 2000; Young et al. 2002). An occultation from 2017 suggests that the increase has peaked (Marques Oliveira et al. 2018; Person et al. 2018).

*Charon*: Prior to the New Horizons flyby, limits on Charon's atmosphere were set to be < 50 nbar (1-σ) from stellar occultation (Gulbis et al. 2006), and the New Horizons UV occultation set limits of < 1.4 pbar (Stern et al. 2017).

Since 2009, stellar occultations have been successfully used to study other TNOs, and the search for an atmospheric signature was carried for a few of them. Most of the largest bodies have already observed during a stellar occultation event and no atmosphere around a TNO has been found so far (other than Pluto and ex-TNO Triton). Upper limits on the presence of putative atmospheres were obtained and they are summarized below and in Table 2.

*Eris*: A stellar occultation of a V = 17.1 star by Eris was observed on November 06, 2010 (Sicardy et al. 2011). Eris, the second biggest TNO after Pluto, was at 95.7 AU and had a surface temperature estimated at 30 K. With methane and nitrogen detected on its surface, it was a good candidate to have retained volatiles. The occultation allowed a constraint on the presence of an isothermal $N_2$ atmosphere to an upper limit of 1 nbar at the surface, with similar limits for $CH_4$ and argon. The occultation revealed that Eris is one of the brightest objects of the solar system, with a geometrical albedo of $p_v$= 0.96 (+0.09/-0.04), which may be caused by a collapsed atmosphere, as discussed above. As the sub-solar temperature of Eris can reach 35 K, the authors mentioned that a local sub-solar atmosphere of $N_2$ may exist, but it would freeze to undetectable values at the limb. It is suggested that, due to the eccentric orbit of Eris, when it approaches its perihelion at 37.8 AU, it may develop a global atmosphere of 2 $\mu$bar.

*Haumea*: An observation of a stellar occultation by Haumea on the February 21, 2017 revealed that it also possesses a dense ring of 70 km in width (Ortiz et al., 2017), making it as the second small solar system object, after (10199) Chariklo (Braga-Ribas et al., 2014a), known to have a ring. Upper limits (1-σ) were obtained for the atmosphere of $P_{surf}$ < 10 nbar ($CH_4$) or $P_{surf}$ < 3 nbar ($N_2$). This is consistent with the lack of spectral evidence of $CH_4$ or $N_2$ on Haumea's surface. Haumea's ring, two moons, and the Haumea dynamical family suggest that volatiles were lost via a disruptive collision (e.g., Schaller & Brown 2007a).

*Makemake*: A Makemake occultation of a V = 18.5 star on April 23, 2011 was detected from seven telescopes in Chile (Ortiz et al., 2012). The preferred solution for the projected shape is an ellipsoid with axes of 1430±9 km and 1502±45 km. Constraining the shape to a circular cross section gives a radius of 1430±9 km, and is also consistent with the observations. The preferred (elliptical) solution gives a geometric albedo at V of $p_V$ = 0.77±0.03. This is a bright albedo, and is consistent with surface freshening by a periodic atmosphere. No global atmosphere was detected, with limits on a global atmosphere of < 12 nbar (1-σ). This is somewhat surprising, as Makemake was a candidate for a global atmosphere, especially given the report of $N_2$ ice from $CH_4$ band shifts. The conclusion is that Makemake has lost almost all of its $N_2$ gas, or that the atmosphere near the limb has frozen to undetectable values (e.g. for a pole-on orientation or a slow rotation/low thermal inertia). Ortiz et al. (2012) also modeled the effect of a local atmosphere on the occultation light curves, motivated by some points close to centrality with an elevated flux a few σ above the noise, and placed loose limits on a local atmosphere of $P_{surf}$ <30 $\mu$bar.

*Quaoar*: A large TNO ($r_0$ = 550 km) with $CH_4$ reported on its surface, Quaoar was a candidate to possess a thin atmosphere. It was observed crossing in front of a star on May 4, 2011, from which upper limits on the presence of an atmosphere were derived (Braga-Ribas et al., 2013). Being at a distance of 42.4 AU, with a geometric albedo of 0.109, a surface equilibrium temperature of 40 K is expected. Considering a $CH_4$ dominant atmosphere, reaching 102 K above 10 km from the surface, an upper limit of $P_{surf}$ < 21 nbar (1-σ) was obtained. As $N_2$ would have a vapor pressure of 66-176 microbar at 40-42 K, the data rule out this scenario.

## INSERT TABLE 2 HERE

*Smaller TNOS*: Additional TNOs have been observed during stellar occultations, and no clear signals of any atmosphere were detected. For these events, there are no published upper limits, because the decrease in stellar flux was abrupt (rather than dimming by atmospheric refraction), and the data did not have high enough quality (cadence or signal-to-noise ratio) for any constraints on a thin atmosphere. The larger bodies (equivalent radii $r_0$ > 200 km) so probed include: Varuna ($r_0$ = 568 km; Sicardy et al. 2010), 2003 $VS_2$ ($r_0$ = 274 km; Benedetti-Rossi et al. 2019), 2002 $TC_{302}$ ($r_0$ = 499 km; Santos-Sanz et al. 2017), 2003 $AZ_{84}$ ($r_0$ = 382 km; Dias-Oliveira et al. 2017) and 2007 $UK_{126}$ (319 km; Benedetti-Rossi et al. 2016).

## 7. Future research

The above discussion suggests some critical directions for future research.

**Surface Compositions**:
- The spectra of Pluto and Triton changes on the timescale of decades, and is related to both changing viewing geometry and possibly real changes in the volatile distribution or physical state of volatile ices (e.g., grain size). The changing compositions of Triton (visual magnitude $V$=13.4) and Pluto ($V$=14.2) can and should continue to be tracked with moderate ground-based telescopes, to decouple these effects.
- For other large TNOs, new understanding of the surface stratification of $CH_4$ and $N_2$ on Eris ($V$=18.7) needs higher spectral resolution and SNR than previous spectra, as does the confirmation of, or constraints on, $CH_4$ on Quaoar (V=18.7), 2007 $OR_{10}$ (V=21.3), Sedna ($V$=20.8), Varuna ($V$=20.1), and Orcus ($V$=19.2). Some progress is likely to come from the new generation of infrared detectors (e.g., NIRES on Keck, or NIHTS on DCT). JWST's spectral sensitivity from 1-5 microns will be very powerful for measuring the volatiles and other species on the surfaces of these and other TNOs (Parker et al. 2016). Both the Thirty Meter Telescope and the European Extremely Large Telescope have planned near-infrared integral field spectrometers as first light instruments, and their various white papers emphasize how they will revolutionize TNO spectroscopy.

**Laboratory work:**
- The vapor pressures of CO and $CH_4$ should be measured at TNO temperatures of 20-40 K, and the vapor pressure of $N_2$ should be remeasured in a modern lab. Also, recent work on the vapor pressures of mixed volatile ices (Trafton 2015; Tan & Kargle 2018) should be tested under laboratory conditions.
- Measurements of the spectrum of $CH_4$ ice in the visible, both pure and in solution with $N_2$, is needed for quantitative analysis of Makemake's $CH_4$ features 0.54-0.62 $\mu$m.

**Atmospheric retention and escape:**
- The most critical work is a better model of why the escape rate at Pluto in 2015 (~5 × $10^{22}$ $N_2$ s$^{-1}$ and ~6 × $10^{25}$ $CH_4$ s$^{-1}$) is so much lower than the energy-limited rate expected pre-encounter (~2 × $10^{27}$ $N_2$ s$^{-1}$), and how that model can be applied to Pluto at other seasons, and to other TNOs.
- Current DSMC work on volatile retention only models a single species, but the observed mixed surfaces with both $N_2$ and $CH_4$ suggest new DSMC work with multiple species. Similarly, local atmospheres (e.g., dome-like atmospheres over the subsolar point) can be more completely modeled.

**Seasonal transport:**
- Models such as Hofgartner et al. (2018) can be used to investigate the transition between global and collisional atmospheres.

**Atmospheric searches:**
- Higher SNR occultations with higher cadence can help the sensitivity, and can help break ambiguity in the cases of suspected local atmospheres. Accurate star positions are now given by the Gaia catalogue, but ephemeris errors are nearly always much greater than the object's apparent angular size. Astrometric positions are needed for the targeted TNOs. One source of positions can be the big and deep surveys like the Large Synoptic Sky Survey that is expected to begin observations in 2022, and will help to update TNOs ephemerides and so improve stellar occultation predictions (Camargo et al. 2018).

## Acknowledgments

Leslie Young was supported in part by the NASA OPR grant NNX14AO45G "Connecting Present and Past KBOs." Felipe Braga Ribas acknowledges CNPq grant 309578/2017-5 and the funding from the European Research Council under the European Communitys H2020 (2014-2020/ERC Grant Agreement no. 669416 "LUCKY STAR"). Apurva Oza was invaluable in the calculations behind Figure 4. Thanks to Bryan Holler for discussions on Section 2, and Larry Trafton for discussion relating to the requirements for a global atmosphere. Mike Brown graciously allowed the reproduction of his figure from Brown et al. (2011), which provided some level of coherence and clarity to our subsequent figures. The paper was greatly improved by a thorough review by Emmanuel Lellouch.

**Table 1. Volatile ices reported or suggested on Transneptunian Objects**

| Body | $CH_4$ ice | $N_2$ ice | CO ice |
|---|---|---|---|
| Triton | Multiple $CH_4$ bands directly detected | $N_2$ band directly detected. $N_2$-rich ice inferred from $CH_4$ band shifts. | Multiple CO bands directly detected |
| (134340) Pluto | Multiple $CH_4$ bands directly detected | $N_2$ band directly detected. $N_2$-rich ice inferred from $CH_4$ band shifts | Multiple CO bands directly detected |
| (136199) Eris | Multiple $CH_4$ bands directly detected | $N_2$-rich ice inferred from $CH_4$ band shifts | No spectral evidence. |
| (136472) Makemake | Multiple $CH_4$ bands directly detected | $N_2$-rich ice inferred from $CH_4$ band shifts | No spectral evidence. |
| (50000) Quaoar | Multiple $CH_4$ bands possibly directly detected | $N_2$ band possibly directly detected. $N_2$-rich ice inferred from $CH_4$ band shifts | No spectral evidence. |
| (225088) 2007 $OR_{10}$ | $CH_4$ inferred by analogy with Quaoar. | No spectral evidence. | No spectral evidence. |
| (90377) Sedna | Weak evidence longward of 2.2 $\mu$m; more work is needed. | Weak evidence reported. | No spectral evidence. |
| (20000) Varuna | Weak evidence at 2.3 $\mu$m; more work is needed. | No spectral evidence. | No spectral evidence. |
| (90482) Orcus | Possible absorption at 2.2 $\mu$m. | No spectral evidence. | No spectral evidence. |

**Table 2. Results on atmospheres from stellar occultations**

| Name | Equivalent radius (km) | Surface Pressure | Reference |
|---|---|---|---|
| Triton | 1350 | 3 - 11 μbar | Elliot et al. 2000 |
| Pluto | 1188 | 4 - 12.7 μbar | Sicardy et al. 2016 |
| Eris | 1163 | < 1 nbar[a] | Sicardy et al. 2011 |
| Haumea | 816 | < 10 nbar[a] | Ortiz et al. 2017 |
| Makemake | 715 | < 12 nbar[a] | Ortiz et al. 2012 |
| Charon | 606 | < 1.4 nbar[a] | Stern et al. 2017 |
| Quaoar | 555 | < 21 nbar[a] | Braga-Ribas et al. 2013 |
| Sedna | 445 | Inconclusive | Braga-Ribas et al. 2014b |

[a] 1-σ upper limits.

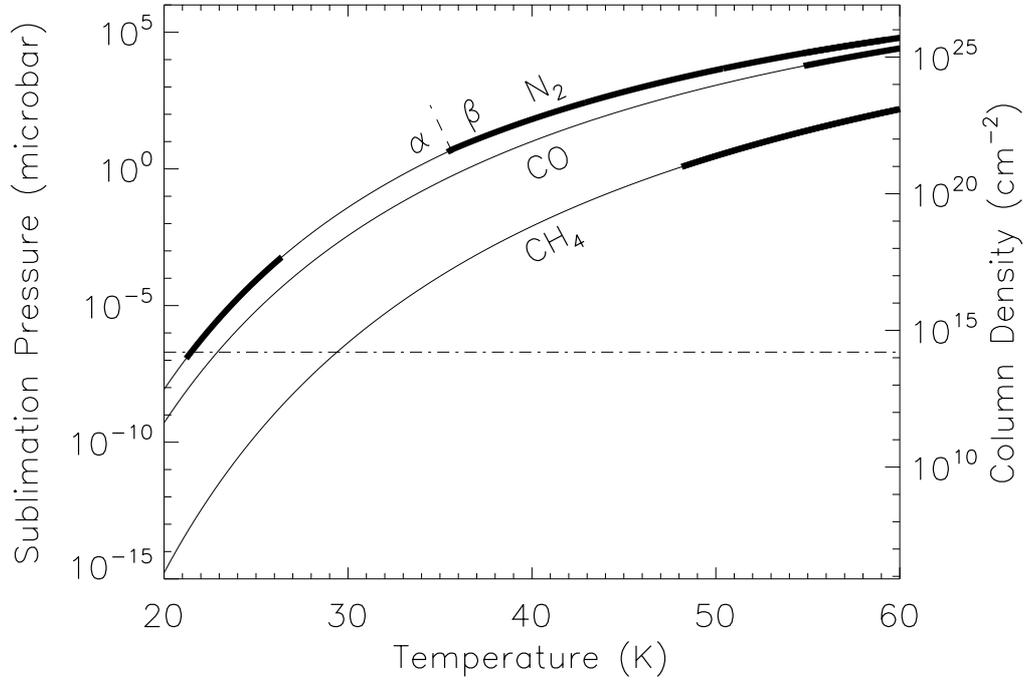

Fig. 1. Sublimation pressure (left axis) for $N_2$, CO, and $CH_4$ (Fray & Schmitt 2009; FS09). The column density (right axis) is calculated for a body with bulk density $\rho$ of 1.9 g cm$^{-3}$ and a surface radius $r_0$ of 500 km for $N_2$ or CO ($\mu = 28$) For a given pressure, the column density scales as $1/(\mu\,\rho\,r_0)$. Thus, the plotted column density (right axis) also applies for $\rho = 1.9$ g cm$^{-3}$ and a surface radius $r_0$ of 875 km for $CH_4$ ($\mu = 16$). Thick lines show the ranges of temperatures at or below 60 K from laboratory measurements included in FS09: for $N_2$, 21.20-26.40 K (Borovik et al. 1960), 35.40-59.17 K (Frels et al. 1974), and 54.78–61.70 K (Giauque & Clayton 1933); for CO, 54.78–68.07 (Shinoda 1969); and for $CH_4$, 48.15–77.65 K (Tickner & Lossing 1951) and 53.15–90.66 K (Armstrong et al. 1955). The α-β phase transition for $N_2$ is indicated (35.61 K). The dot-dashed line indicates the rough transition between ballistic and collisional atmospheres, for $\rho = 1.9$ g cm$^{-3}$ and $r_0 = 500$ km.

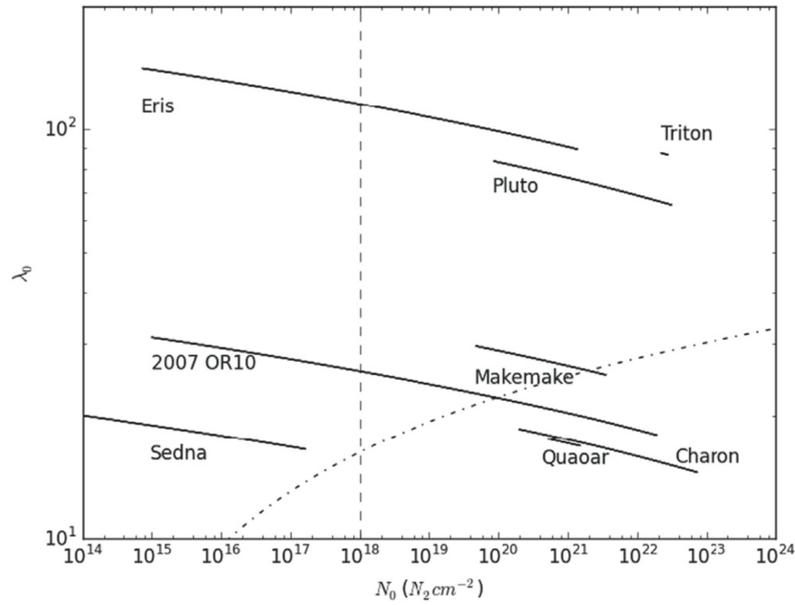

Fig. 2. Range of surface Jeans parameter, $\lambda_0$, vs. total column density, $N_0$, over the orbits of a number of TNOs assuming an $N_2$ atmosphere (Johnson et al. 2015) using $\varepsilon = 0.8$ and $A=0.67$, appropriate for an early TNO with a bright surface. For Makemake, 2007 $OR_{10}$, and Sedna we used $\rho = 1.8$ g cm$^{-3}$. Sedna's values are shown only near perihelion as its aphelion ($N_0 \ll 10^{14}$ cm$^{-2}$) is off-scale. Dashed line: $N_0 = 10^{18}$ $N_2$ cm$^{-2}$; to the right the $CH_4$ and $N_2$ components are sufficient so that escape is mainly driven by solar heating of the atmosphere. Dotted–dashed line: surface-heating-induced escape rate is smaller than Jeans formula below this line (warmer surface or lower surface gravity) and greater above this line. From Johnson et al. 2015. Reproduced by permission of the AAS.

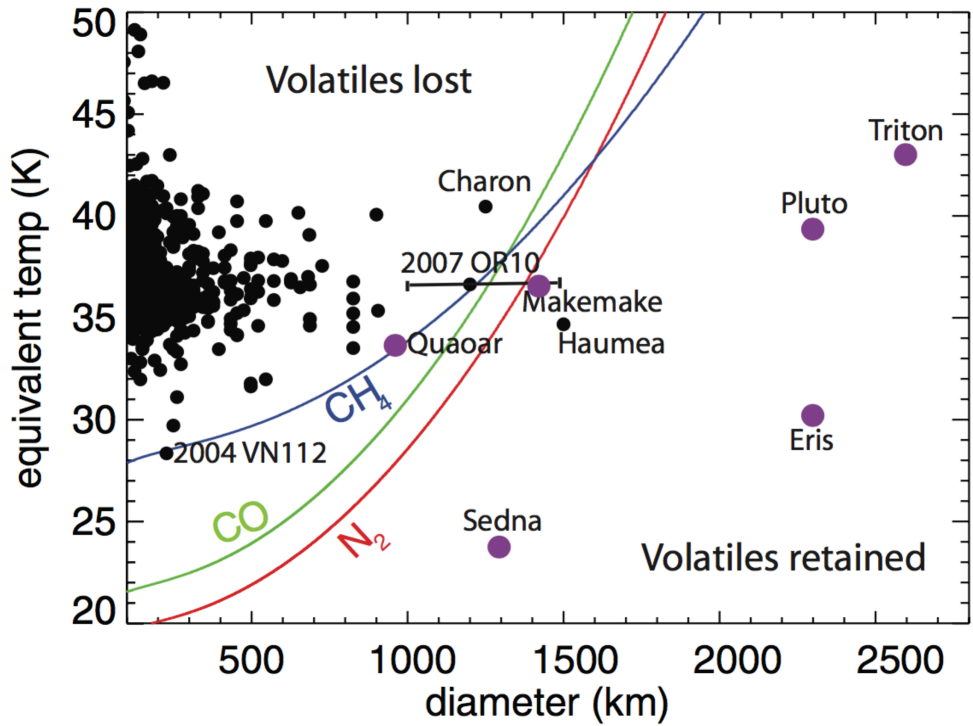

Figure 3. Plot of volatile retention and loss in the Kuiper belt, from Brown et al. 2011, updated from Schaller & Brown (2007b). Reproduced by permission of the AAS. Using the surface-Jeans formulation for atmospheric escape, objects to the left of the $CH_4$, CO, and $N_2$ lines lose surface volatiles over the age of the solar system.

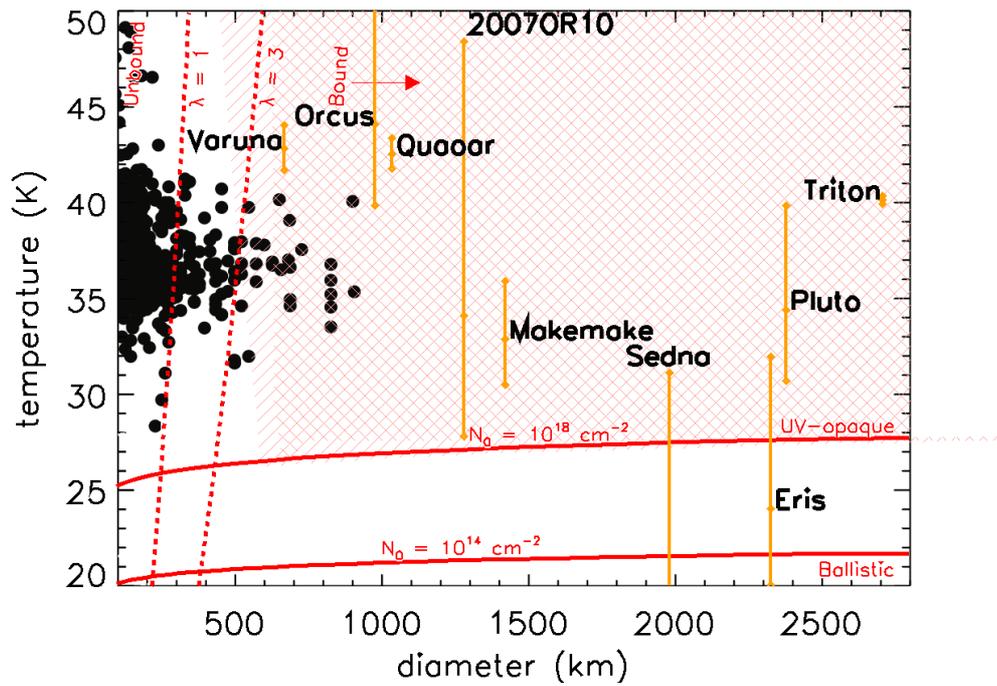

Figure 4. Diagram of escape regimes. Yellow: The nine bodies in Table 1 have been plotted: solid vertical lines show the variation in the average equilibrium temperatures between aphelion and perihelion (Eq. 1), using diameters and albedos from the *tnosarecool* database (that is, updated from Brown et al. 2011 and Johnson et al. 2015). Black: Diameters and temperatures for other bodies are taken from the retention plot of Brown et al. (2011), as plotted in Fig 3.. Solid red: Lines of constant column density at the surface for two critical values: for $N_0 = 10^{14}$ cm$^{-3}$ defining the classical surface-bounded exobase, and $N_0 = 10^{18}$ cm$^{-3}$ for an $N_2$ atmosphere with 3% $CH_4$ to be opaque to Ly-$\alpha$. Dashed red: Lines of constant Jeans parameter for two critical values dividing the unbound and bound atmosphere (Volkov et al. 2011b). Hashed red: Atmospheric heating dominates the loss rate in the shaded region (adopted from the shaded regions in Fig 3 of Johnson et al. 2015) for $A = 0.1$ (single hash) and $A = 0.67$ (cross-hatched). Outside the hashed area, the surface-heated escape rate for warm, small bodies is much smaller than the surface-Jeans estimate, but the surface-heated escape rate for cold, large bodies is slightly larger than the surface-Jeans estimate (cf. dot-dashed line in Fig 1).

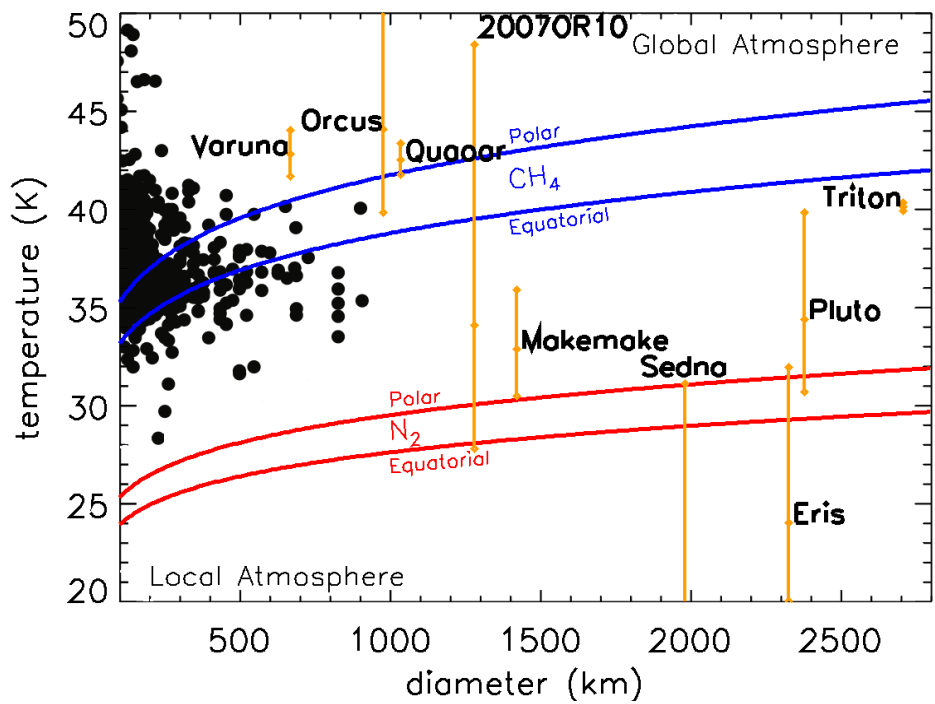

Figure 5. In a global atmosphere, pressures are high enough so that the sublimation winds can keep surface pressures and volatile ice temperatures nearly uniform over the globe. Minimum temperatures for a global atmosphere are plotted, assuming a surface uniformly covered with $CH_4$ ice (blue) or $N_2$ ice (red), for two extreme sub-solar latitudes. Several bodies would have global atmospheres at some portion of their orbit if covered with $N_2$ ice (as reported on Triton, Pluto, Eris, Makemake, and possibly Quaoar). A much smaller number of bodies would have global atmospheres at some portion of their orbit if covered with only $CH_4$ ice (as reported possibly on Varuna, Orcus, Quaoar, Sedna, and 2007 OR10). Thermal inertia and non-uniform volatile ice coverage can change the range of temperatures actually achieved. Temperatures and diameters are as in Fig. 4.

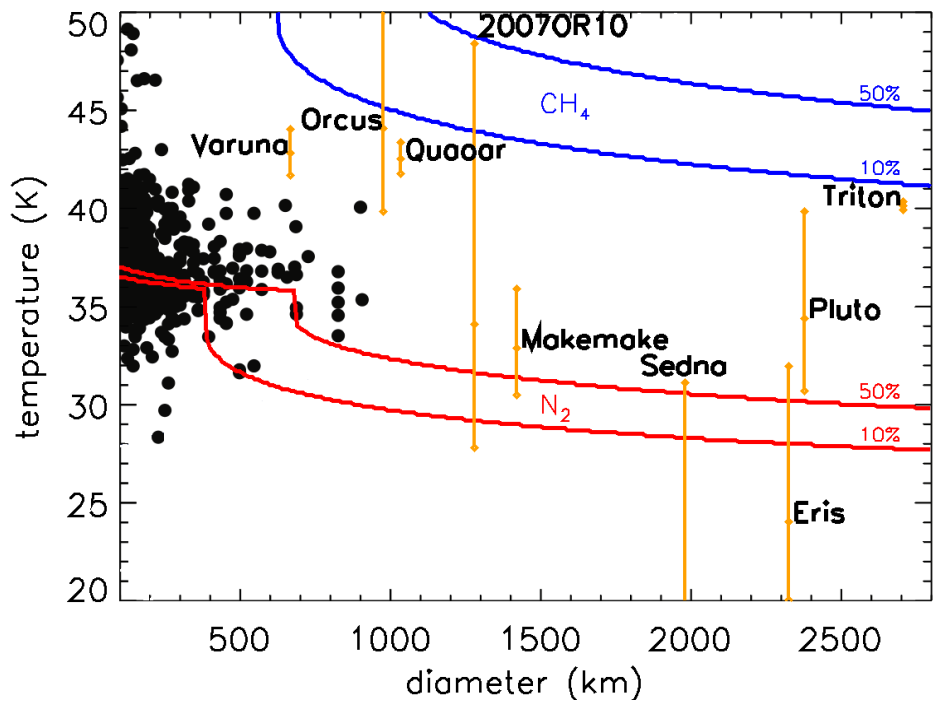

Figure 6. Detectability of TNO atmospheres by ground-based stellar occultation, under the assumption that the atmosphere is isothermal, and both atmosphere and surface are at the plotted temperature. The blue ($CH_4$) and red ($N_2$) detectability limits are for 50% or 10% drop in stellar flux for a surface-grazing ray. Near 35 K for $N_2$ and 50 K for $CH_4$, the surface-grazing ray causes a central flash (Eq 10, second term), which leads to a discontinuity in the derived limits; the plotted limits are somewhat conservative near this discontinuity. Temperatures and diameters are as in Fig. 4.

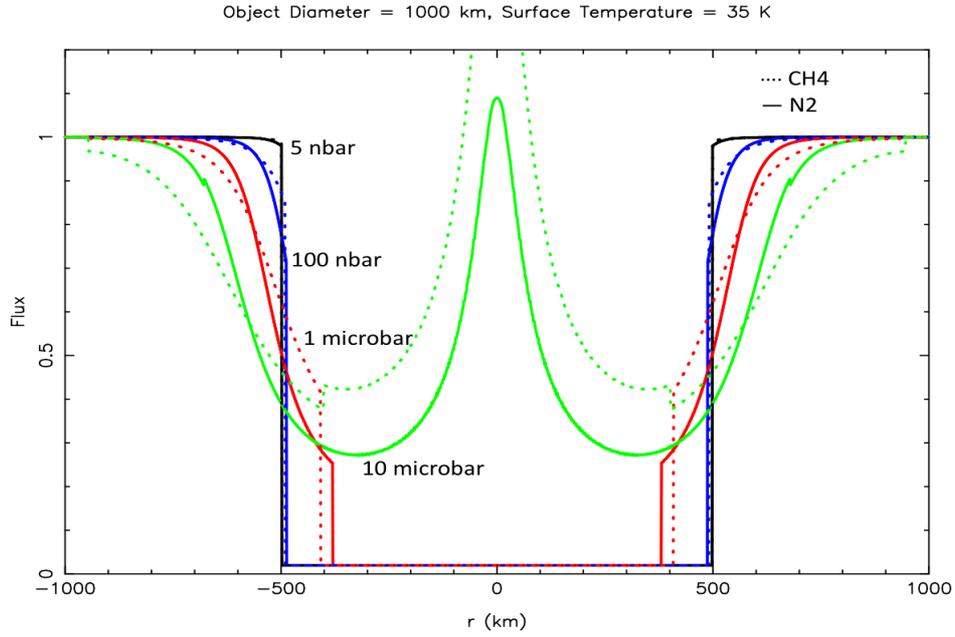

Figure 7: Synthetic light curve models of normalized stellar flux vs. shadow radius for an object with $r_0 = 500$ km, density 1.9 g cm$^{-3}$, $\Delta = 40$ AU, with a surface temperature of $T_0 = 35$ K and a 5 K km-1 gradient to a 100 K upper atmosphere. Continuous lines are for an N$_2$-dominated atmosphere with surface pressures and column densities of $p_0 = 5$ nbar, $N_0 = 4.0 \times 10^{18}$ molecule cm$^{-2}$ (black); 100 nbar, $8.1 \times 10^{19}$ molecule cm$^{-2}$ (blue); 1 $\mu$bar, $8.1 \times 10^{20}$ molecule cm$^{-2}$ (red), and 10 $\mu$bar, $8.1 \times 10^{21}$ molecule cm$^{-2}$ (green). Dashed lines are for a CH$_4$-dominated atmosphere with surface pressures and column densities of 0.005 $\mu$bar, $7.1 \times 10^{18}$ molecule cm$^{-2}$ (black); 0.1 $\mu$bar, $1.4 \times 10^{20}$ molecule cm$^{-2}$ (blue); 1 $\mu$bar, $1.4 \times 10^{21}$ molecule cm$^{-2}$ (red); and 10 $\mu$bar, $1.4 \times 10^{22}$ molecule cm$^{-2}$ (green).